# Visualizing Heavy Fermion Confinement and Pauli-Limited Superconductivity in Layered CeCoIn$_5$

András Gyenis[1,†,‡], Benjamin E. Feldman[1,†,§], Mallika T. Randeria[1,†], Gabriel A. Peterson[1,¶], Eric D. Bauer[2], Pegor Aynajian[3], and Ali Yazdani[1,*]

[1]Joseph Henry Laboratories of Physics, Department of Physics, Princeton University, Princeton, NJ 08544

[2]Los Alamos National Laboratory, Los Alamos, NM 87545

[3]Department of Physics, Applied Physics and Astronomy, Binghamton University, Binghamton, NY 13902

[‡]Present address: Department of Electrical Engineering, Princeton University, Princeton, NJ 08544

[§]Present address: Department of Physics, Stanford University, Stanford, CA 94305

[¶]Present address: National Institute of Standards and Technology, Boulder, CO 80305

[†]These authors contributed equally to this work.

*Corresponding author. Email: yazdani@princeton.edu

## Abstract

Layered material structures play a key role in enhancing electron-electron interactions to create correlated metallic phases that can transform into unconventional superconducting states. The quasi-two-dimensional electronic properties of such compounds are often inferred indirectly through examination of their bulk properties. Here we use scanning tunneling microscopy and spectroscopy to directly probe in cross section the quasi-two-dimensional correlated electronic states of the heavy fermion superconductor CeCoIn$_5$. Our measurements reveal the strong confined nature of heavy quasi-particles, anisotropy of tunneling characteristics, and layer-by-layer modulated behavior of the precursor pseudogap gap phase in this compound. Examining the interlayer coupled superconducting state at low temperatures, we find that the orientation of line defects relative to the *d*-wave order parameter determines whether in-gap states form due to scattering. Spectroscopic imaging of the anisotropic magnetic vortex cores directly characterizes the short interlayer superconducting coherence length and shows an electronic phase separation near the upper critical in-plane magnetic field, consistent with a Pauli-limited first-order phase transition into a pseudogap phase.

*Introduction*

A central theme of the research on unconventional superconductivity has been its strong relationship to reduced dimensionality (*1-4*). For example, the layered crystal structure of high-$T_c$ superconductors gives rise to strongly two-dimensional (2D) electronic behavior, which increases the many-body correlation effects that are an essential ingredient for unconventional superconductivity. The heavy fermion superconductor CeCoIn$_5$, which has many similarities to the high-$T_c$ cuprates (*5-10*), also has a layered crystal structure built up from the heavy fermion antiferromagnet CeIn$_3$ (*11*) separated by CoIn$_2$ stacks. Bulk measurements of CeCoIn$_5$ show signatures of an anisotropic, quasi-2D electronic structure (*12-17*), but in contrast to the cuprates, there are also contributions from 3D bands that result in a smaller electronic anisotropy (*18*). Among the Ce-based heavy fermion compounds, CeCoIn$_5$ has the highest transition temperature at ambient pressure, which correlates with its electronic dimensionality as illustrated by isovalent substitutions (*19-21*) and layer engineering (*22-23*). Like the cuprates, superconductivity in CeCoIn$_5$ has a $d_{x^2-y^2}$ symmetry (*24-31*) and there are indications of a pseudogap phase (*24,30,32-35*) as well as other ordered phases that compete or coexist with superconductivity, such as the spin-density wave order identified as the *Q*-phase (*36-38*). This phase appears at high magnetic fields, just before the upper critical field associated with a Pauli limited transition into the pseudogap phase (*30,39-40*).

Here we introduce a new experimental approach to investigate the electronic structure of CeCoIn$_5$: we use a scanning tunneling microscope (STM) to study its properties in cross section. Our measurements directly probe the layer dependence of the electronic states, and represent the first cross-sectional study of a layered superconducting system. Our approach reveals important features of the correlated quasi-2D electronic structure in CeCoIn$_5$, including confinement of heavy quasi-particles on the atomic scale and the layer dependence of its pseudogap. Furthermore, in the superconducting state, the cross-sectional geometry enables us to probe the direction dependent response of the $d_{x^2-y^2}$ order parameter to scattering from defects as well as spatially resolve the nature of the vortex phase and its first order Paul limited phase transition into the pseudogap state.

*Results*

To probe the quasi-2D nature of electronic behavior in the normal and superconducting phases of CeCoIn$_5$, we cleave samples along the [100] orientation (parallel to the *b-c* surface) *in situ* in an ultra-high vacuum STM. Based on the crystal structure, we expect that the resulting surfaces expose a cross-sectional cut of the quasi-2D layers of this compound (Fig. 1a). The crystal structure also suggests that the surface termination in the [100] orientation will be either a Ce-Co-In$_2$ layer or an In$_3$ layer, and STM topographical images indeed show two different surfaces for the cleaved samples in the *b-c* plane (Fig. 1b). One is atomically ordered and smooth, and we label it surface S; the other appears more disordered, and we refer to it as surface R. We identify surface R as the In$_3$ layer and attribute the quasi-ordered bumps in the STM images to surface reconstruction [see the details in the Supplementary Information (SI)]. We assign the atomically ordered surface S to the Ce-Co-In$_2$ layer (Fig. 1c), which is expected to be offset from surface R with a step height of half of the lattice constant in the *a-b* plane, as found experimentally (Fig. 1d). The morphology of the surfaces allows us to identify the position of the quasi-2D layers [in previous studies (*30,41-42*) referred as layer A and B] based on topographic images (SI). Our measurements of surface S reveal an anisotropic atomic lattice consistent with the crystal structure of CeCoIn$_5$, and we focus on high-resolution measurements of this surface in the remainder of this work.

Our ability to access CeCoIn$_5$ layers in cross section provides a unique opportunity to address the role of the layered structure of this compound on its electronic properties. Previous STM studies (*30,41-42*) of CeCoIn$_5$ on samples cleaved along the [001] orientation had been used to examine the layer dependence of the electronic properties by studying multiple surfaces perpendicular to the *c*-axis that were terminated with different layers. In those experiments, it was found that the STM tip couples differently to the heavy or light quasi-particles depending on the surface atomic termination, resulting in changes in the tunneling spectra. When tunneling is sensitive to the light *spd* electrons (on layer A), the spectra show a hybridization gap for such quasi-particles, whereas on layer B a stronger coupling to the *f*-electrons yields an asymmetric double peak in the spectra associated with the formation of the dispersing heavy *f*-band. In the current work, spectroscopic measurements on the exposed *b-c* plane enable us to investigate the composite nature of electronic states and their variation in the different quasi-2D layers while studying a single atomic surface (Fig. 1a). Measurements on surface S at *T* = 10 K show two types of spectra depending on the atomic positioning of the tip: one corresponding to layer A (Fig. 2a), in which a hybridization gap is observed and the other to layer B (Fig. 2b), where a double peak feature is resolved. One type of spectrum evolves smoothly into the other as the STM tip examines the layers of CeCoIn$_5$ in cross section (Fig. 2c). This smooth progression illustrates the remarkable property that the observed electron mass varies significantly on the atomic scale within a single (100) unit cell and that it is strictly associated with the 2D layers. The transition between light and heavy nature of the excitations can be captured by a simple model (Fig. 2d) that considers the spatial dependence of the tunneling sensitivity (SI).

One intriguing observation is that the spectroscopic signatures of the *spd* electron hybridization depends on whether the tunneling occurs perpendicular or parallel to the 2D layers. Previous measurements performed on the *a-b* plane (*30-31,41*) indicate the presence of two gap-like features in the tunneling spectra (with energy scales of around 40 meV and 15 meV). These can be associated with a direction dependent hybridization gap [or gaps (*43-44*)] based on quasi-particle interference (QPI) measurements (*41*). In contrast, in our current cross sectional experiments, we only observe one feature that matches the smaller hybridization gap when tunneling perpendicular to the same layer (Fig. 2a). Our data show that in addition to the previously observed in-plane anisotropy of the measured hybridization gap, the geometry of the STM measurement strongly influences the sensitivity of such measurements.

Mapping variations of the local density of states (LDOS) in the tunneling spectra on the *b-c* surface provides evidence for strong confinement of quasi-particles within the quasi-2D layers of CeCoIn$_5$. Figure 3a shows a region where several islands of surface R act as scattering potentials, giving rise to modulations in the LDOS from QPI (*45*) (Fig. 3b). Far from the defects (for example at the bottom left corner of Fig. 3a-b) the QPI signal is absent and the LDOS exhibits a periodic modulation along the *c* axis. This is the same behavior as observed in Fig. 2, and it further demonstrates that the stacked quasi-2D layers have different electronic character. Near the islands, our cross-sectional imaging geometry reveals a preferential direction for quasi-particle scattering: the interference waves are oriented along the *b* axis, whereas the modulation is almost absent in the direction of the *c* axis. This strongly confined scattering behavior can be further demonstrated by taking a Fourier transform of the conductance map (Fig. 3d), which reveals three significant scattering vectors. The $Q_1$ vector with the strongest intensity and the weaker $Q_2$ are in the [010] direction and correspond to scattering along the quasi-2D layers (along the *b* axis). The presence of 3D bands in CeCoIn$_5$ leads to a scattering vector $Q_3$, which has both [010] and [001] components [with $Q_3 \approx (0, 0.37, 0.69)$ r.l.u.], although this scattering signal is substantially weaker than the in-plane signal at $Q_1$. We note that no scattering vector can be detected purely in the [001] direction, which indicates the low probability of electrons moving perpendicular to the quasi-2D layers (in the direction of the *c* axis).

High-resolution conductance mapping (Fig. 3c) illustrates an additional aspect of the confinement of the quasiparticles: the strength of the QPI signal is strongly suppressed on lines on top of layer B. This is also visible in Fig. 3e, which displays the QPI modulation as a function of distance from the island on two neighboring atomic planes (one is on top of layer A and the other one is on top of layer B). On layer A, the interference signal exhibits a decaying, periodic, long-wavelength modulation of about 30 Å, whereas it is almost absent on layer B, showing that the dominant scattering vector is mainly detectable on layer A. Energy-resolved QPI measurements along the [010] direction (see SI for details) reveal two major bands (a heavy and a light band) and shows that the long-wavelength signal ($Q_1$) at this energy arises from scattering involving light bands, so its prominence on layer A is consistent with our discussion above and previous STM measurements (*30-31,41-42*).

Next, we study the superconducting state of CeCoIn$_5$, where our cross-sectional geometry allows us to map the antinodal direction of the $d_{x^2-y^2}$ order parameter, which points out of plane from the exposed surface (Fig. 4a inset). When the sample is cooled to $T \approx 400$ mK, well below its transition temperature, the spectrum exhibits a sharp, superconducting gap at the Fermi energy (Fig. 4a), which is unchanged as the STM tip crosses the quasi-2D layers (Fig. 4c), reflecting the fact that the coherence length is much longer than the interlayer spacing. The gap size $\Delta_{SC} \approx 550$ µeV is similar to previously measured values (*25,28,30-31*). Tunneling into a layered *d*-wave superconductor in cross section has not been previously demonstrated; our measurements offer a new approach for studying its response to impurities. Examining the spatial variation of the gap in the *b-c* surface, we find no variation in $\Delta_{SC}$ across atomic step edges (Fig. 4e-g), in stark contrast to a previous measurement of scattering events in the *a-b* plane (*30*). In that experiment, suppression of the superconducting gap was observed due to the sign change of the order parameter for electrons and holes with different in-plane momenta. In our geometry, we find that the gap is insensitive to such defects, which is consistent with the *b-c* surface of CeCoIn$_5$ having a *d*-wave order parameter with a uniform phase (see schematic in Fig. 4g).

Application of a magnetic field induces vortices and eventually quenches superconductivity through a first order phase transition to create a pseudogap state in CeCoIn$_5$ (*24,30,32-35*). We first discuss our STM spectroscopic measurements which reveal signatures of superconductivity up to a magnetic field $H^*$, which is higher than the upper critical field $H_{c2}$ obtained from bulk thermodynamic studies (36). The evolution of the spectra with magnetic field (measured between vortices, see below) is shown in Fig. 4b. There is a jump in the zero-bias conductance between 12.3 T and 12.5 T, which is associated with a first order transition, in this case out of the superconducting state into a pseudogap state. Similar jumps in the spectra were reported in a previous study for the field applied along the $c$-axis (30). However, this $H^*$ = 12.3 T transition field is above the bulk $H_{c2}$ = 11.8 T measured with thermodynamic techniques in CeCoIn$_5$ samples from the same batch. Differences between measurements of $H_{c2}$ from transport and thermodynamic studies have been previously reported in related heavy fermion systems [SI and (46-48)]. While we currently do not have a full explanation for this apparent difference between the STM-measured $H^*$ and the bulk $H_{c2}$ values, our STM data suggest that superconductivity survives locally to fields larger than the bulk $H_{c2}$.

Unlike the superconducting state, the pseudogap phase of CeCoIn$_5$ shows a layer dependent behavior similar to the confined electronic nature of the normal state discussed above. The LDOS exhibits pronounced variations on the atomic scale, as shown in Fig. 4d for $H$ = 13 T (which is above the bulk $H_{c2}$ and surface measured $H^*$). In layer A, the spectrum resembles the normal state at zero field and displays only the hybridization gap; in contrast, layer B exhibits an additional suppression of conductance over a smaller energy range around the Fermi level, indicative of a pseudogap. These results are consistent with previous observations that the pseudogap in CeCoIn$_5$ is observed only when tunneling into the layer B, where there is strong coupling to $f$ electrons (30). By imaging in cross section, we not only confirm that the pseudogap feature is associated with the layers exhibiting heavy electronic behavior but also demonstrate that this phase varies on the atomic scale on a single cleaved surface, in sharp contrast to the superconducting phase. Observation of a spectroscopic signature of a pseudogap is consistent with transport studies of CeCoIn$_5$ (24,32,34-35), although there has been effort to explain this observation based on a heavy quasi-particle band structure effect (49).

In the presence of a magnetic field, the superconducting state develops vortices, and our cross sectional technique allows us to visualize the anisotropy of the electronic behavior in the resulting vortex state of CeCoIn$_5$. By probing vortices in the $b$-$c$ plane, we extract a direction dependent characteristic coherence length, map the unusual vortex lattice structure, and directly image the transition of a Pauli limited superconductor. A series of maps obtained in the same area between 9 T and 12.3 T are shown on Figs. 5a-e, where the lighter elongated regions of high conductance correspond to vortex cores, and the red dots represent the fitted center of mass of each vortex. We present background subtracted conductance maps to suppress the effect of conductance variations due to different surface terminations and defects in the field of view [SI and (50)]. Although the shape of individual vortices is disordered due to surface inhomogeneity and impurities, they exhibit an overall ellipsoid shape. To suppress the effects of inhomogeneity, we overlay all (~90) vortices measured at various fields through their center of mass. The resulting average vortex displays an azimuthally asymmetric core (Fig. 5f), which is a manifestation of the anisotropic coherence length in the *b-c* plane. Although a detailed model calculation of the local density of states that includes the multiband nature of CeCoIn$_5$ is needed to fully characterize the vortex core shape, we extract characteristic lengths from our data by fitting the decay of the vortex conductance as function of distance *r* from the center at different angles $\phi$ with respect to the *c* axis according to $G(r,\phi) \sim e^{-r/\xi(\phi)}$ (51). From this fit, we find characteristic lengths of $\xi^c = 30$ Å along the *c* axis to $\xi^b = 65$ Å along the *b* axis (Fig. 5g), which are consistent with values estimated for the in- and out-of-plane coherence lengths from measurements of the angle dependence of $H_{c2}$ (14). Conductance maps taken at various energies confirm the presence of the zero bias peak inside the vortex core (Fig. 5h).

Our high magnetic field measurements in the *b-c* plane demonstrate an unusual structural transition in the vortex lattice which is different from the ones found when the magnetic field was applied in the [001] or [110] direction (52-54). As illustrated in Fig. 5, for *H* < 11 T, the vortices are arranged in a distorted hexagonal Abrikosov lattice with a field-independent β = 41 ± 2° opening angle, in excellent agreement with small angle neutron scattering studies (54). However, when the magnetic field is increased above 11 T, a previously unreported vortex lattice transition occurs. In this phase, the vortices are arranged in rows along the *c* direction, with larger spacing along the *a*-axis. One possible cause of such a change of the vortex lattice could be the onset of the Q-phase. However, such transition in the vortex lattice could also result from various effects such as the strong local anisotropy of the vortices, nonlocal electrodynamic effects between them, or superconducting gap symmetry effects (55).

Finally, by mapping the electronic structure in close proximity of *H\** we directly image the transition of a Pauli limited superconductor to its normal state (39-40,53). Generally, two effects of the applied magnetic field govern the physics of a superconducting condensate: the kinetic energy of the supercurrent around the vortices and the Pauli energy of the electron spins coupled to the external field. In an orbital limited superconductor the superconductivity is suppressed by the overlap of vortices, while in a Pauli limited case, the Cooper pairs are destroyed by breaking the spin-singlet state, as is the case in CeCoIn$_5$. Imaging the vortex state near the critical field at *H\** = 12.3 T (Fig. 5e) shows the coexistence of a normal region and vortices in the same field of view, while above $H^*$ only normal regions are present (SI). Due to the short anisotropic coherence length, the distance between the cores and the orbiting supercurrents is large, which allows the Pauli paramagnetic effects to dominate the orbital effects in CeCoIn$_5$. Moreover, the emergence of domains is expected to occur for first order phase transitions; the coexistence of both normal and superconducting regions therefore provides a direct visualization of the first order superconducting phase transition in CeCoIn$_5$.

*Discussion*

In conclusion, we have explored the influence of the layered material structure and reduced effective dimensionality of CeCoIn$_5$ on its confined electronic properties by utilizing the STM as a cross sectional probe for samples cleaved along the [100] direction. Spectroscopic measurements performed in the normal and superconducting states demonstrate the effects of quasi-two-dimensionality, from varying effective electron mass on the atomic scale and confined quasiparticle scattering to layer dependent pseudogap behavior and anisotropic vortex structure in the superconducting state. Imaging these dramatic effects in cross section offers a direct illustration of quasi-2D electronic behavior in this archetypal correlated electron system.

## Methods

The single-crystal samples used for the measurements were grown from excess indium at Los Alamos National Laboratory. Crystals with large thickness in the *c* direction were chosen for the measurements, cut into suitable sizes (with dimensions in all directions of ~0.5-2 mm), oriented and glued to the sample holder with the (100) surface facing up. An aluminium post with the same horizontal dimension was glued to the top of the sample and used to cleave the sample along the *c*-axis in ultra-high vacuum at room temperature. Immediately after cleaving the samples, they were inserted into our home-built STMs. We used a variable temperature STM for the T = 10 - 20 K temperature measurements and a dilution fridge STM for the low temperature ($T \approx 400$ mK) and high magnetic field experiments. A large number of samples (around 30) were cleaved in both setups, and each cleaved sample was approached multiple times (using long range piezoelectric motion). On the cleaved samples, we found atomically flat surfaces suitable for STM measurements with a success rate around 10% of the approaches. Differential conductance measurements were performed using standard lock-in techniques, with voltage bias applied to the sample.

## Acknowledgements

We thank Brian B. Zhou, Eduardo H. da Silva Neto, and Shashank Misra for the fruitful discussions. Work at Princeton was primarily supported by DOE-BES (DE-FG02-07ER46419) and Gordon and Betty Moore Foundation as part of EPiQS initiative (GBMF4530). This project was also made possible using the facilities at Princeton Nanoscale Microscopy Laboratory supported by grants through NSF-MRSEC programs through the Princeton Center for Complex Materials DMR-1420541, NSF-DMR -1608848 , ONR-N00014-14-1-0330, ONR-N00014-13-10661, and Eric and Wendy Schmidt Transformative Technology Fund at Princeton. Work at Los Alamos National Laboratory was performed under the auspices of the US Department of Energy, Office of Basic Energy Sciences, Division of Materials Sciences and Engineering. B.E.F. acknowledges support from the Dicke Fellowship. M.T.R. acknowledges support from the NSF Graduate Research Fellowship. P.A. acknowledges funding from the U.S. National Science Foundation CAREER under award No. NSF-DMR 1654482.

## Author contributions

A. G., B. E. F., M. T. R. and G. A. P. performed the STM measurements. A. G., B. E. F., M. T. R. and P. A. analyzed the data. E. D. B. synthesized and characterized the materials. A. G., B. E. F., M. T. R., P. A., and A. Y. wrote the manuscript. All authors commented on the manuscript.

The authors declare no competing financial interests.

Figure captions

**Figure 1 STM topographic images of the (100) surface of CeCoIn$_5$.** a, Schematic diagram of the bulk crystal structure of CeCoIn$_5$ showing the two possible surface terminations (S and R) when cleaving along the [100] orientation. Lines indicate the positions of layers A, B and C. The lattice constants are $a = b = 4.6$ Å and $c = 7.52$ Å. b, Constant current topographic images ($V_{bias} = -100$ mV, $I_{setpoint} = 1.2$ nA) of the (100) surface morphology, which displays a large atomically ordered surface S and small islands of the reconstructed surface R. c, Topographic image of a few unit cell area on surface S with red rectangular showing a unit cell on the *b-c* plane. d, Topographic linecut along the white line indicated on panel b, which shows the height difference between surface S and R and corresponds to $a / 2 = 2.3$ Å.

**Figure 2 Atomic scale variation of the fermion mass.** a-b, STM tunneling spectra ($V_{bias} = -100$ meV and $I_{setpoint} = 1.7$ nA) acquired along a $b = 4.6$ Å long line parallel to the *b* axis, which display negligible spatial variation and corresponds to light (layer A) and heavy mass (layer B), respectively. c, Tunneling spectra across a line parallel to the *c* axis between two consecutive B layers showing alternating peak-dip structure and indicating that the observed electron mass varies with the position in the unit cell. Color-coded dots on the top panels show the position of the measured spectrum on the surface. The same smooth background is subtracted from all spectra and the curves are vertically shifted for clarity. d, Calculated tunneling spectra along the *c* axis. Inset shows the $t_f/t_c$ ratio (see SI for details).

**Figure 3 Quasi-particle interference on the (100) surface.** a, Topographic image of surface S where the conductance map was acquired ($V_{bias} = -70$ mV, $I_{setpoint} = 1$ nA). Inset shows an enlarged topographic image with the position of layer A and B indicated. b, Conductance map at $E = -70$ meV energy showing quasiparticle standing waves around the atomic islands. The conductance of the islands is artificially saturated for clarity. c, Enlarged conductance map, which demonstrates the strongly one-dimensional scattering of the quasiparticles. Arrows indicate the position of layer A and B. d, Symmetrized Fourier transform of the conductance maps shown in b. Green rectangle shows the border of the unit cell in reciprocal space. e, The modulation of the LDOS along a line parallel to *b* axis (shown as white line on panel a) on top of layer B (blue) and top of layer A (green). Dark yellow curve shows the exponential decay envelope of the interference pattern obtained by fitting the data with $G(d) = G_0 \sin\left(\frac{2\pi}{\lambda_{\text{QPI}}} d + \varphi\right) e^{-d/\xi_{\text{QPI}}} + G_{\text{mean}}$, where *d* is the distance from the island, $\lambda_{\text{QPI}} = 31$ Å is the wavelength of the quasi-particle signal, $\xi_{\text{QPI}} = 52.4$ Å is the decay length, $\varphi$ is the phase of the signal and $G_{\text{mean}}$ is the mean conductance.

**Figure 4 Superconductivity and pseudogap phase in (100) CeCoIn$_5$.** a, Averaged tunneling spectra ($V_{bias} = -30$ meV, $I_{setpoint} = 1$ nA) obtained in the superconducting phase at $T = 400$ mK, exhibiting a sharp superconducting gap ($\Delta_{\text{SC}}$) around the Fermi energy. Inset: Schematic picture showing the relative position of the STM tip and the superconducting order parameter. b, Averaged tunneling spectra ($V_{bias} = -20$ meV, $I_{setpoint} = 500$ pA) acquired in high magnetic field around $H^*$ show an abrupt jump of the zero-bias conductance between 12.3 T and 12.5 T. c, High-energy resolved measurement ($V_{bias} = -6$ meV, $I_{setpoint} = 300$ pA) of the superconducting gap along the *c* axis with the color of the curves indicating the position of the spectra on the S surface (green corresponding to the top of layer A and blue to the top of layer B). d, Tunneling conductance measurement ($V_{bias}$ = - 30 meV, $I_{setpoint}$ = 1 nA) along a line on S surface above the superconducting transition at $H$ = 13 T, which reveals a layer-dependent pseudogap ($\Delta_{PG}$) opening around layer B, whereas the spectra on layer A exhibits only the hybridization gap ($\Delta_H$). e, The superconducting gap evolution ($V_{bias}$ = - 10 mV, $I_{setpoint}$ = 500 pA) along a 140 Å long line through a double atomic step edge as indicated on the topographic image in panel f. The superconducting gap is insensitive to the potential variation due to the step edge. g, Schematic picture of the position of the *d*-wave order parameter and the STM tip along the step edge.

**Figure 5 Anisotropic vortices and vortex lattice transition.** a-e, Subtracted conductance maps ($G_{sub}$) obtained on a 500 Å x 500 Å area with magnetic fields applied parallel to the *a* axis, which show elongated vortices on the (100) surface. Red dots indicate the fitted centers of mass of the vortices. Dashed line displays the fit through the centers of mass of the vortices to determine the opening angle β. The colorbar corresponds to the normalized subtracted conductance map $G_{sub,norm} = G_{sub}/|\overline{G_{sub}}|$, where $\overline{G_{sub}}$ is the mean of the subtracted conductance value over the entire field of view. f, Averaged vortex shape obtained by overlaying 90 measured vortices at different fields. $\phi$ corresponds to the angle with respect to the *c* axis. g, Extracted effective coherence length as a function of angle $\phi$. h, Spatially averaged density of states in the vortex core (green), far from the vortex (blue) and their difference (red), which show the existence of the bound states inside the vortex.

Figure 1

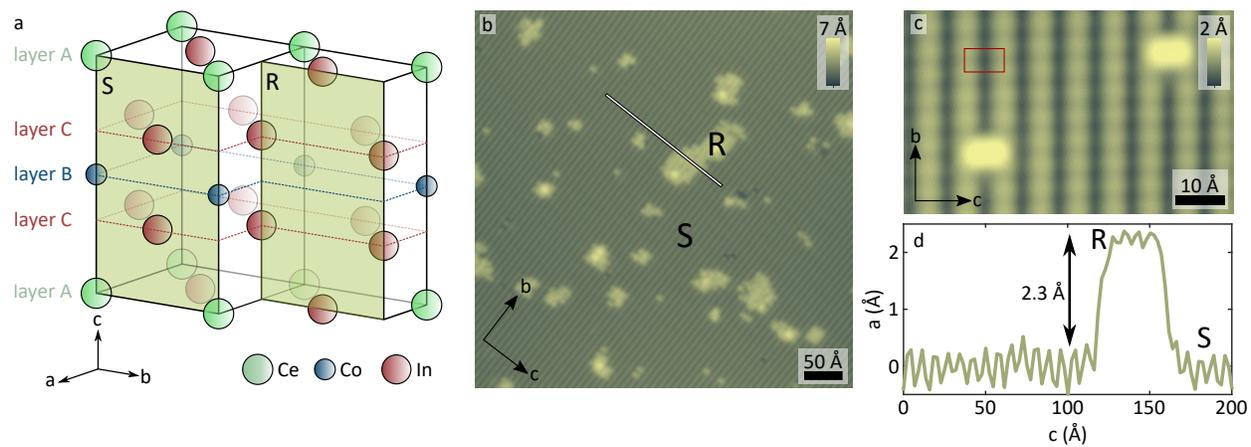

Figure 2

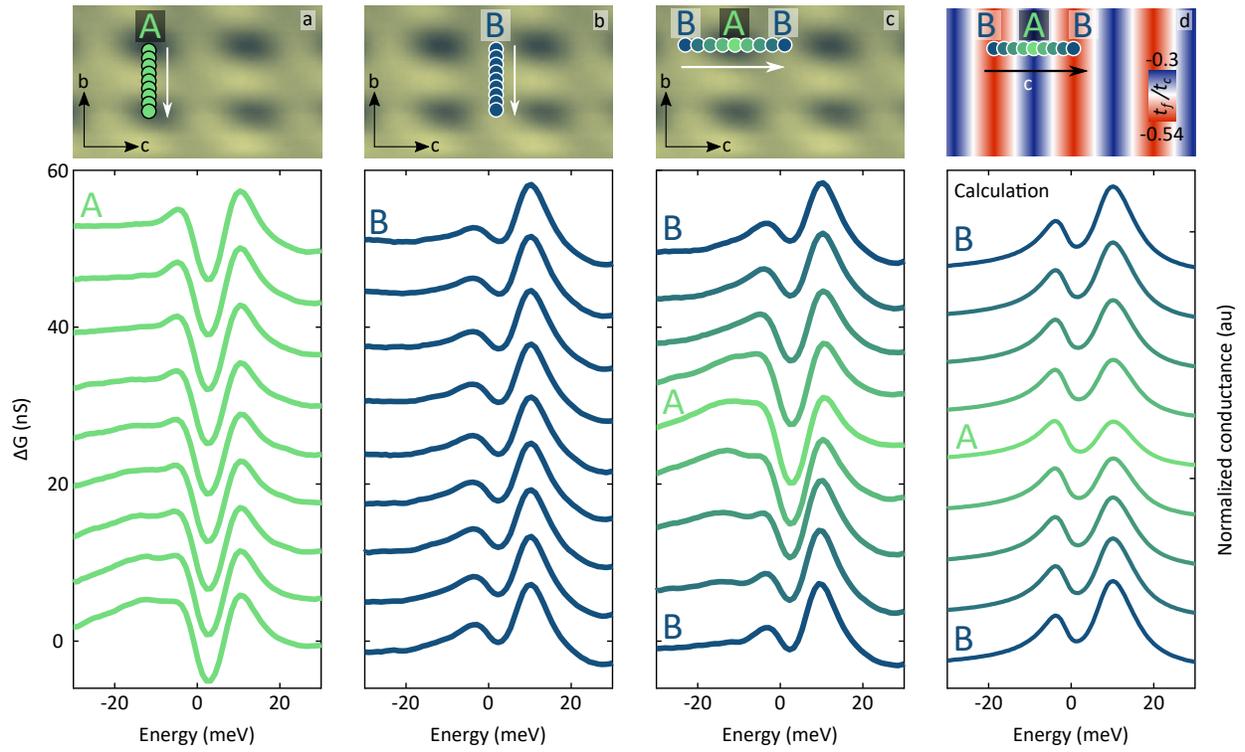

Figure 3

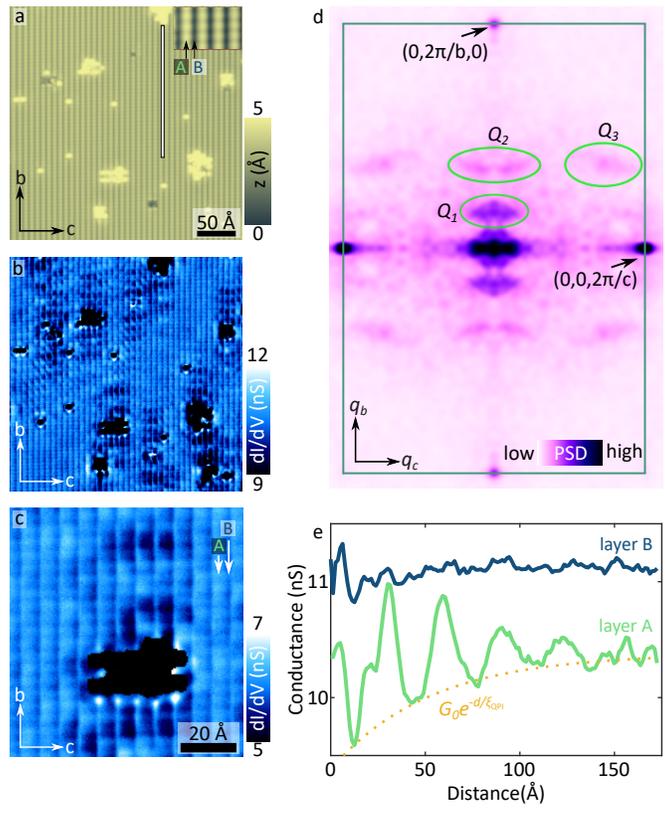

Figure 4

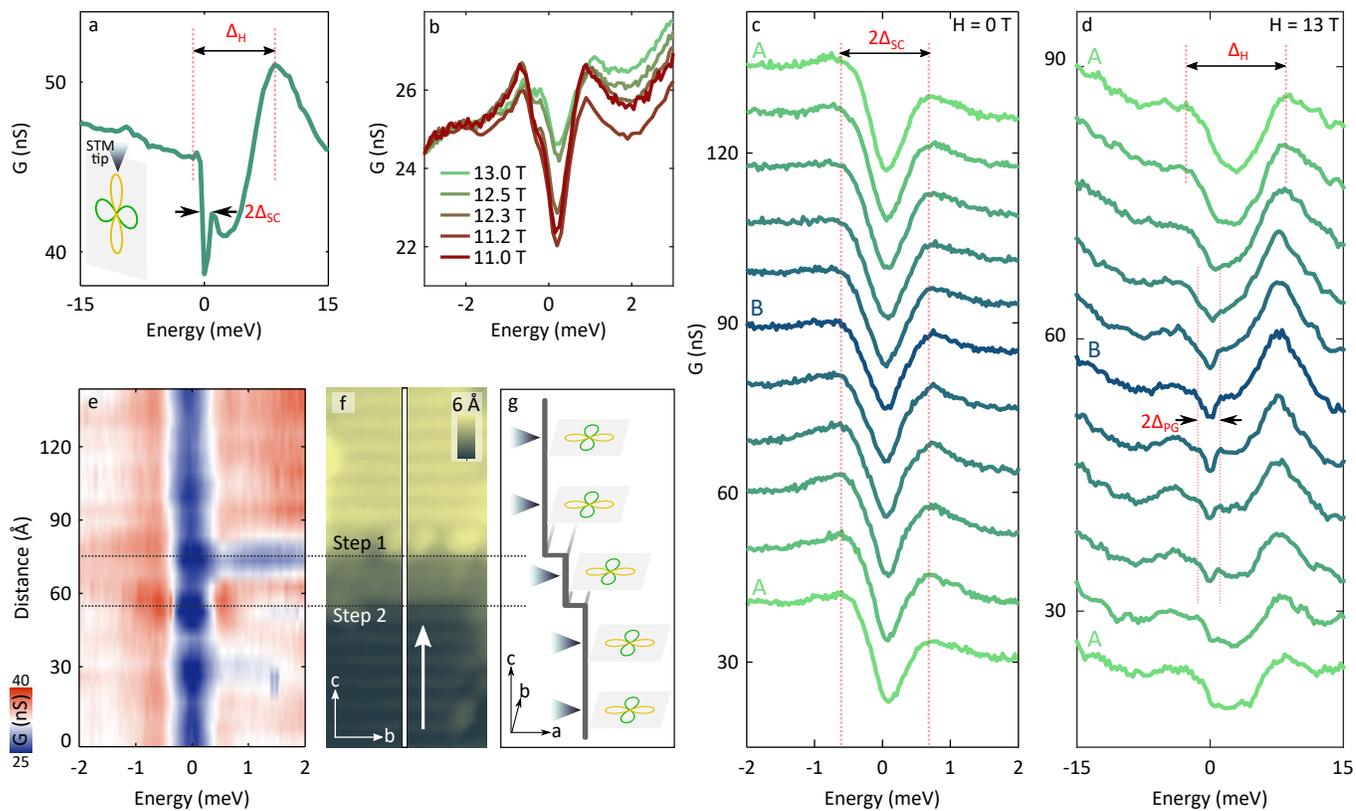

Figure 5

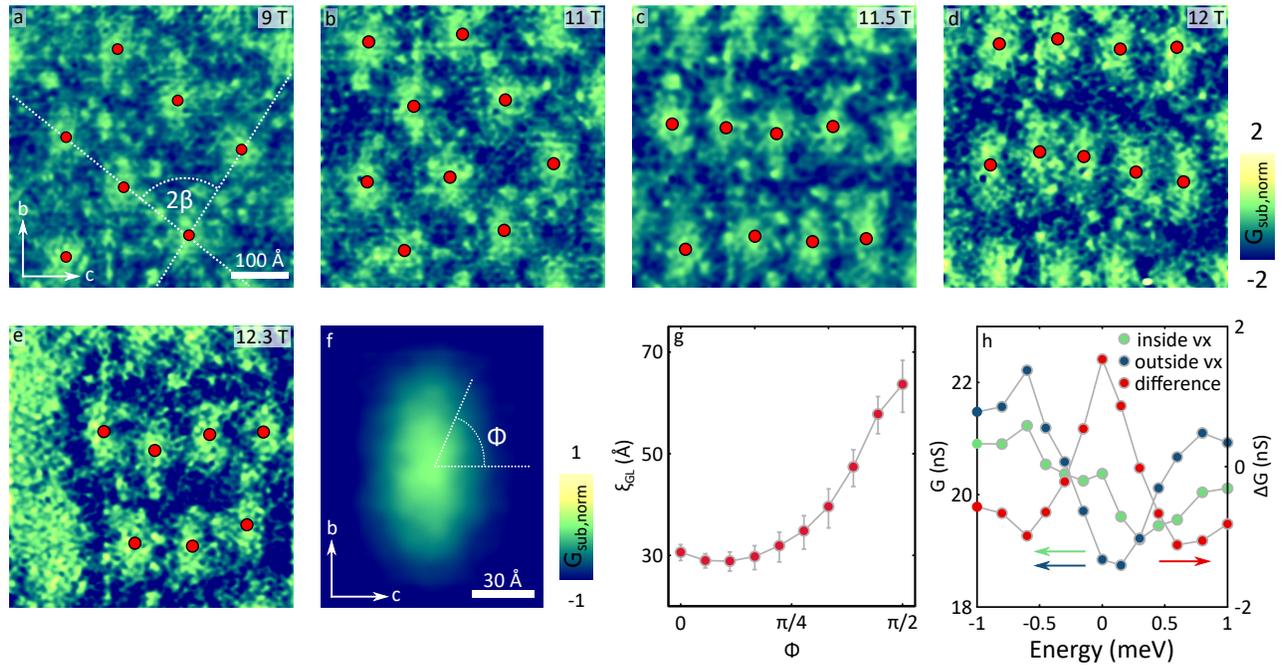

# Supplementary Information for

# Visualizing Heavy Fermion Confinement and Pauli Limited Superconductivity in Layered CeCoIn$_5$

András Gyenis[†], Benjamin E. Feldman[†], Mallika T. Randeria[†], Gabriel A. Peterson, Eric D. Bauer, Pegor Aynajian, and Ali Yazdani[*]

[†]These authors contributed equally to this work.

[*]Corresponding author. Email: yazdani@princeton.edu

**Section I. Observed morphologies of the (100) cleaved surface**

We cleaved several samples along the [100] direction, approached multiple different areas on each cleaved sample, and consistently found two distinct surface morphologies (Fig. S1). For most regions we approached, the topographic images revealed morphology-I (Fig. S1a), whereas in rare cases morphology-II (Fig. S1b) was observed. In both cases, we found two surface terminations, one of which (surface R) is reconstructed (Fig. S1c-d). We associate this surface with the layer containing In atoms (In$_3$ layer) because previous measurements on CeCoIn$_5$ cleaved along the (001) direction (*S1*) also showed a reconstructed In$_2$ surface. Surface S in the second kind of morphology shows a rectangular lattice with lines running in the high symmetry directions (Fig. S1f), whereas in the first kind of morphology it shows a mixture of rectangular (Fig. S1e) and hexagonal order (Fig. S1c). The topography in the hexagonal regions is inverted relative to the reconstructed surface but has identical lattice spacing. We therefore believe that it either reflects the influence of a subsurface In$_3$ layer or results from the detailed atomic arrangement after the cleave. On surface S, not all atoms in the unit cell are visible, but the structural modulation resembles the (100) unit cell, and the position of the two-dimensional layers can be indirectly determined (see Section II). Although the origin of the two different morphologies is unclear, the normal vector of the surfaces, the in-plane atomic modulations, and the step heights are in good agreement with the expected crystal structure of the (100) plane of CeCoIn$_5$. We also note that both types of morphologies have been observed on a single cleaved sample.

**Section II. Origin of the reconstructed In surface and determination of the position of the 2D layers**

The Ce, Co, and In atoms are not all visible on the (100) cleaved surfaces, so we use an indirect method to determine the position of the 2D atomic layers based on the unusual structure of the reconstructed $In_3$ surface. As Fig. S2a shows, two inequivalent In atoms exist on surface R: In(1) (from layer A) and In(2) (from layer C). By examining the quasi-ordered reconstructed surfaces observed in our STM topographic images, we conclude that the simplest model to explain the pattern involves a superstructure containing one In(1) atom and four surrounding In(2) atoms. This assignment matches well to both the size and the orientation of the features on the reconstructed surface of the more common morphology. Figure S2b shows that the superstructure lattice can shift by half of its unit cell in one direction, corresponding to a shift by one unit cell in the *b* direction (Fig. S2b); this may occur spontaneously or in response to a missing In atom on the surface. As a result, we observe superstructures running in the *b* direction with frequent instances of shifts within these lines.

Since the In(1) atom is part of layer A, we determine that layer A is located along the midpoint of the large circular superstructures, whereas layer B is located between them (Fig. S2c). By following these 2D planes through a step edge, we can also identify the layers on surface S (Figs. S2c-d). The assignment of the 2D layers based on the reconstructed surface is consistent with the assignment based on spectroscopic features that we observe on each layer (double peak or gap in the spectrum in Figs. 2a-c).

**Section III. Theoretical model of the tunneling density of states**

To capture the spectroscopic features (Fig. 2), we use a theoretical model (*S2-S3*), which was previously successfully applied to data acquired on the *a-b* surface of CeCoIn$_5$ (*S1*). In this theory, the differential conductance *dI/dV* can be obtained from the interference of tunneling paths into two channels: the light and heavy electronic excitations.

The dispersion of the conduction band is

$$\epsilon_{\mathbf{k}} = 2t(\cos k_x + \cos k_y) - \mu,$$

whereas for the heavy band it is

$$\chi_k = -2\chi_0(\cos k_x + \cos k_y) - 4\chi_1 \cos k_x \cos k_y + \epsilon_f,$$

where $t$ is the nearest neighbor hopping, $\mu$ is the chemical potential, $\chi_0$ and $\chi_1$ correspond to the antiferromagnetic correlation between the $f$ moments and $\epsilon_f$ can be associated with the chemical potential for the $f$ electrons.

The components of the full Green's function are

$$G_{ff}(\mathbf{k},\omega) = \left\{[G_{ff}^0(\mathbf{k},\omega)]^{-1} - s^2 G_{cc}^0(\mathbf{k},\omega)\right\}^{-1},$$

$$G_{cc}(\mathbf{k},\omega) = \left\{[G_{cc}^0(\mathbf{k},\omega)]^{-1} - s^2 G_{ff}^0(\mathbf{k},\omega)\right\}^{-1},$$

$$G_{cf}(\mathbf{k},\omega) = -G_{cc}^0(\mathbf{k},\omega) s G_{ff}(\mathbf{k},\omega),$$

where $s$ describes the coupling between the magnetic moments and the conduction electrons and $G_{ff}^0(\mathbf{k},\omega) = (\omega - \chi_\mathbf{k} + i\Gamma_f)^{-1}$, $G_{cc}^0(\mathbf{k},\omega) = (\omega - \varepsilon_\mathbf{k} + i\Gamma_c)^{-1}$ with the corresponding inverse lifetimes of $\Gamma_f$ and $\Gamma_c$.

The $dI/dV$ spectrum can be approximated as

$$\frac{dI(\mathbf{r},\omega)}{dV} \propto -\mathrm{Im} \sum_{i,j=1}^{2} \left[\hat{t}\hat{G}(\mathbf{r},\omega)\hat{t}\right]_{ij}$$

where $\hat{t} = \begin{bmatrix} t_c & 0 \\ 0 & t_f \end{bmatrix}$ describes the sensitivity to tunnel into heavy or light part of the electrons.

In our calculation, we use $t = 200$ meV, $\mu = 2t$, $\chi_0 = 0.01t$, $\chi_1 = 0.06\chi_0$, $\epsilon_f = 0.035t$, $s = 0.15t$, and $\Gamma_c = \Gamma_f = 0.015t$, and vary the $t_f/t_c$ ratio as a function of position with respect to the two-dimensional layers (Fig. 2d).

**Section IV. Energy-momentum dispersion and lifetime of the quasiparticle interference signal**

In Fig. S3, we show the amplitude of the Fourier transform of differential conductance as a function of energy and momentum in the [010] direction. The data illustrate that $Q_1$ is slowly dispersing ($v \approx 0.8$ eVÅ) at energies far from the Fermi level and can be associated with the light part of the hybridized band structure. We also observe a flat band around zero energy corresponding to the enhanced scattering of heavy excitations similar to measurements on CeCoIn$_5$ cleaved along the [001] direction (*S1, S4-S6*). The measured decaying QPI signal (Fig. 3e) and the dispersion relation allow us to extract the lifetime of the quasiparticles. As discussed in the main text, fitting the amplitude of the modulation to an exponential decay function yields a decay length $\xi_{\text{QPI}} = 52$ Å. Based on the measured dispersion relation, this results in a lifetime of $\tau_{\text{QPI}} = \xi_{\text{QPI}}/v \approx 40$ fs. This value is in good agreement with the quasiparticle lifetime obtained from the $\Delta E = 10$ meV width of the spectral function measured on surface B (*S1*), from which $\tau = \hbar/\Delta E \approx 65$ fs.

**Section V. Identifying the upper critical field from STM, transport and thermodynamic studies**

Our STM measurements carried out in magnetic fields applied in the [100] direction show the absence of the signatures of superconductivity at the field of *H\** = 12.3 T, which is higher than the previously reported upper critical field $H_{c2}$ values (*S7*). Our samples are of high quality and show bulk thermodynamic $H_{c2}$ = 11.8 T, consistent with many other previous studies. Here, we discuss possible reasons for the experimental observation that superconductivity locally survives above the bulk $H_{c2}$.

We first emphasize that our STM measurements clearly show that there is a superconducting gap (measured outside of the vortices), which evolves smoothly from lower fields, and survives up to 12.3 T (Fig. 4b). The difference between the superconducting gap and the pseudogap is clear in our measurements, as there is a jump in the zero energy conductance between 12.3 T and 12.5 T. Spectroscopic imaging as a function of field (Fig. 5) also clearly shows the vortex lattice surviving through the bulk $H_{c2}$, and the lack of overlap between the vortices is consistent with the Pauli limited nature of superconductivity in this compound. Furthermore, the observation of coexisting normal regions and superconducting areas with vortices at 12.3 T is consistent with a first order superconducting phase transition.

A second important point is that in the CeMIn$_5$ (M = Co, Rh, Ir) superconductors, one often finds a significant difference between the upper critical field determined from bulk measurements (e.g., specific heat, nuclear magnetic resonance) compared to $H_{c2}$ determined from transport measurements [see, for example, (*S8-S10*)]. Usually, this difference between $H_{c2}^{\text{transport}}$ and $H_{c2}^{\text{bulk}}$ occurs when antiferromagnetism is present above the superconducting transition. CeIrIn$_5$ is a notable exception, with no obvious antiferromagnetic transition observed above the bulk superconducting transition at $T_c^{\text{bulk}}$ = 0.4 K, while $T_c^{\text{transport}}$ = 1.3 K with a corresponding difference in $H_{c2}^{\text{bulk}}$ = 0.9 T < $H_{c2}^{\text{transport}}$ = 7 T for fields applied in the a-b plane (*S10*). Based on the similarities of CeCoIn$_5$ to CeIrIn$_5$ (i.e., no antiferromagnetism present above $T_c$ in zero magnetic field) and on our experimental findings (presence of the vortex lattice and evidence of the superconducting gap from dI/dV measurements) we conclude that superconductivity in CeCoIn$_5$ is observed up to $H^* = 12.3$ T ($H||a$).

Our STM measurements, which are uniquely sensitive to the electronic structure on the surface, are the first local measurements to provide insight into the superconducting properties of CeCoIn$_5$ near $H_{c2}$ for fields applied in the [100] direction. We hope that this result will stimulate further work to understand the origin of the discrepancy in measured upper critical field from different techniques.

**Section VI. Conductance maps in magnetic field**

As we discuss in the main text, we use a background subtraction scheme to enhance the visibility of the vortices. We define the subtracted conductance maps as $G_{sub}(x, y, E, H) = G(x, y, E, H) - G(x, y, E, H_{ref})$, where $G(x, y, E, H)$ is the real space conductance value acquired at energy $E$, magnetic field $H$ at spatial position of $(x, y)$, while $H_{ref}$ corresponds to the reference magnetic field (*S11*). As a reference, we choose conductance maps obtained at $H_{ref} = 13$ T ($H_{ref} > H^*$) instead of zero field because the magnetic field dramatically reduces some of the impurity scattering resonances (*S12*). We note that using $H = 0$ T as reference leads to qualitatively similar results. Finally, we choose $E = 0$ because the sharpest vortex imaging contrast is achieved at the Fermi energy (Fig. 5h and Fig. S6). We use a drift correction code to compensate for the small displacements of the acquired conductance maps at different magnetic fields.

In Fig. S4, we plot the raw conductance maps (Fig. S4b-d) and the subtracted maps (Fig. S4e-f) obtained on the surface shown in Fig. S4a. The maps show that the average conductance varies significantly on the different R and S surfaces. Nonetheless, the subtraction allows us to image the vortices. Choosing the $H = 0$ T map as a reference leads to dark, circular regions associated with impurity scattering resonances (Fig. S4e), which are absent in the subtracted map where $H = 13$ T is the reference field (Fig. S4f).

In addition to the data presented in the main text, we show subtracted conductance maps at other magnetic fields (Fig. S5) and at other energies (Fig. S6).

## Supplementary figure captions

**Figure S1 STM topographic images of the observed surface morphologies of (100) $CeCoIn_5$.** a-b, Constant current topographic images ($V_{bias}$ = - 60 mV, $I_{setpoint}$ = 100 pA and $V_{bias}$ = - 100 mV, $I_{setpoint}$ = 1.2 nA) of the two observed types of surface morphology, which display the consecutive reconstructed surface R and atomically ordered surface S. c-f, Enlarged topographic images of surface R and surface S in the case of the two morphologies.

**Figure S2 Identification of the position of the two-dimensional layers.** a, The In-terminated layer R on the *b-c* surface of $CeCoIn_5$. Due to surface reconstruction, five In atoms (highlighted with red circles) form the circular objects observed in the STM images. The quasi-lattice of the reconstruction has *2b x c* quasi-periodicity. b, When one In atom is absent (*e.g.*, due to the cleaving procedure), the corresponding reconstructed sphere is shifted by a lattice constant in the *b* direction. c, The center of the circular superstructures corresponds to the position of layer A (black arrows), and their edges correspond to layer B (blue arrows). d, Topographic image of R and S surfaces separated by a single step edge, showing the identified layers (blue lines and arrows correspond to layer B, black arrows to layer A). The horizontal green lines indicate the lattice in the [010] direction.

**Figure S3 Energy-momentum structure of the quasi-particle interference.** Fourier transform amplitude of the conductance maps along the [010] direction obtained at different energies shows two pronounced features. At large negative energies (from -10 meV to -80 meV), the $Q_1$ vector (around 0.16 r.l.u.) slowly disperses, which indicates that it originates from the light conduction band. Around the Fermi energy, a rapidly dispersing signal appears, which is the result of scattering between the heavy bands.

**Figure S4 Comparison of subtracted conductance maps.** a, Topographic image of the *b-c* surface of CeCoIn$_5$, where all the presented vortex maps were obtained. b-d, Conductance maps at various magnetic fields. e-f, Subtracted conductance maps using $H$ = 0 T and $H$ = 13 T, respectively, as reference fields.

**Figure S5 Vortices on the *b-c* surface of CeCoIn$_5$.** Subtracted conductance maps ($G_{sub}$) at various fields in the vicinity of the upper critical field ($V_{bias}$ = - 10 mV, $I_{setpoint}$ = 300 pA; reference field: $H$ = 13 T).

**Figure S6 Energy structure of the vortices.** a-g, Subtracted conductance maps at different energies showing that the vortices are most visible at zero energy. The red and white squares show the spatial positions where the averaged conductance values were obtained at each energy for inside and outside the vortex core, respectively (Fig. 5h).

# Figure S1

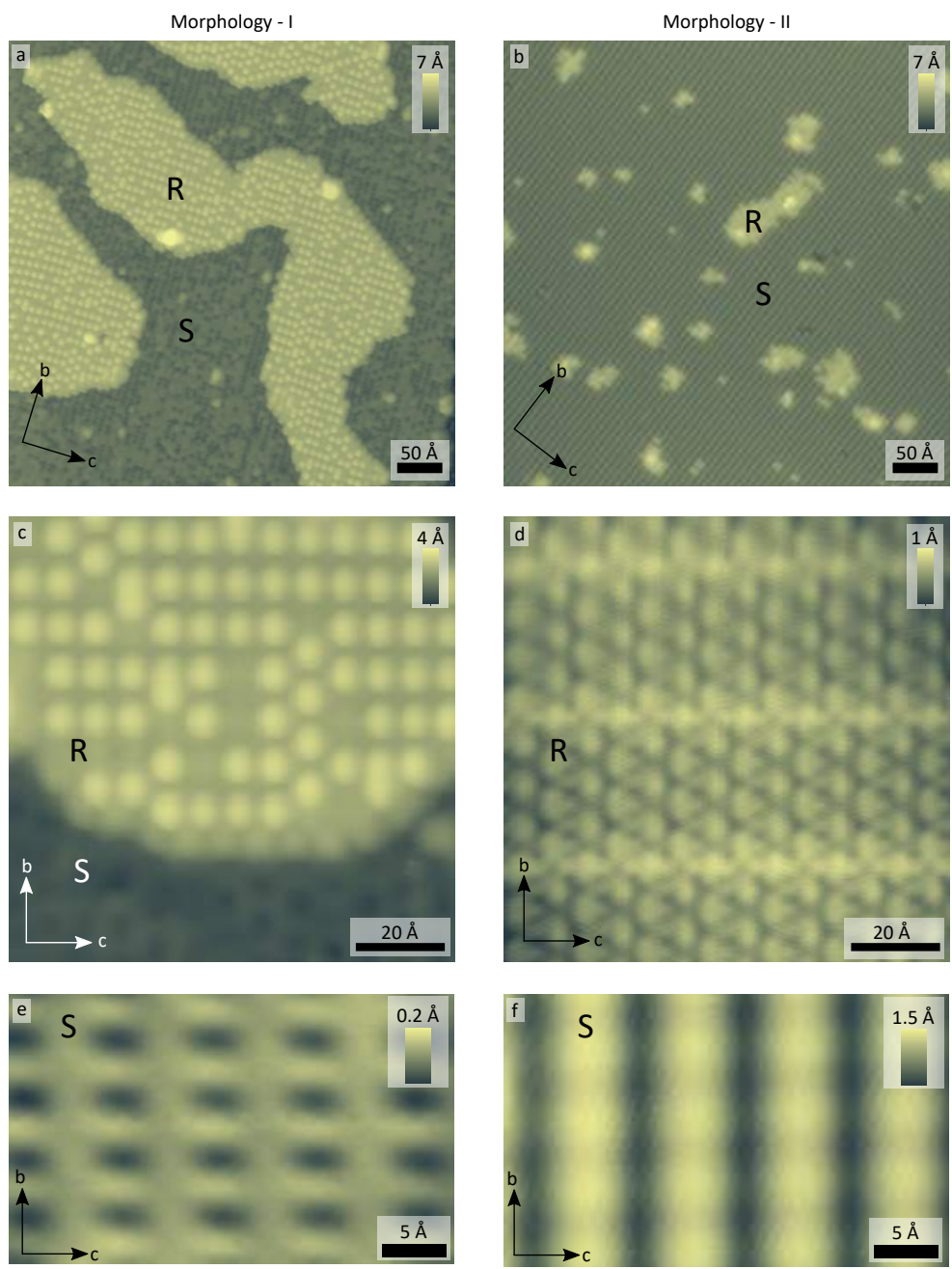

# Figure S2

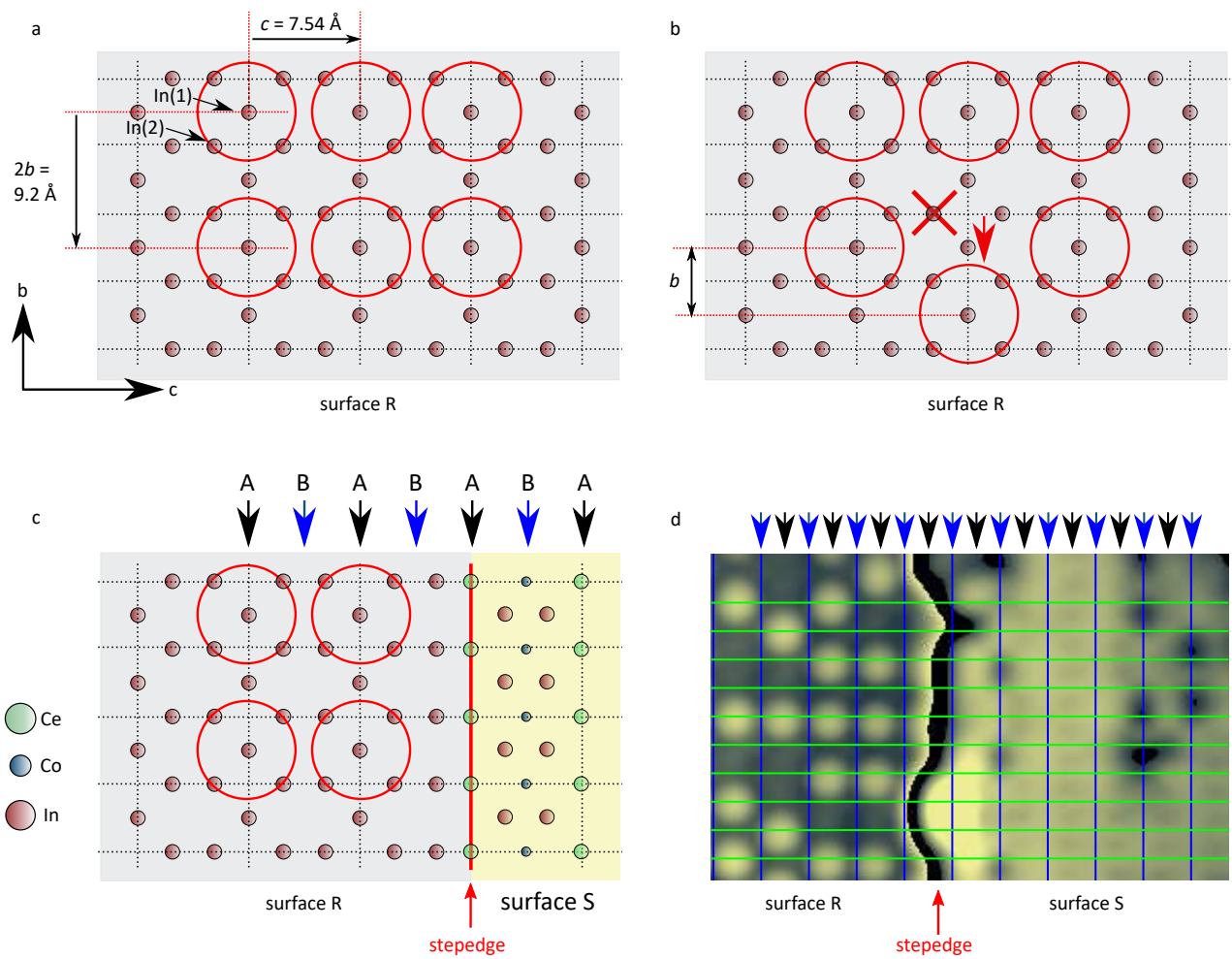

# Figure S3

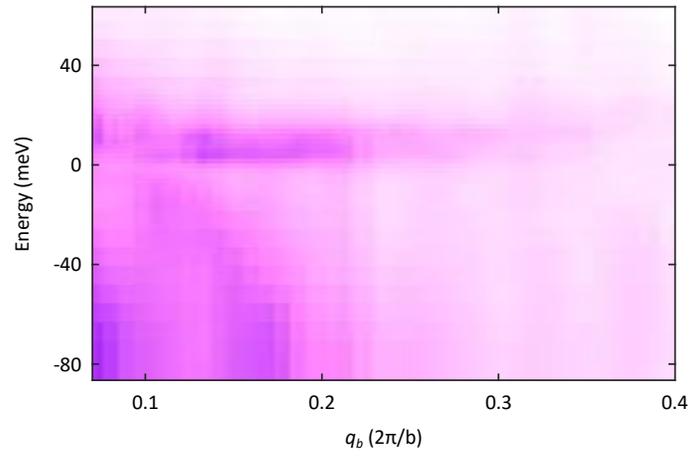

# Figure S4

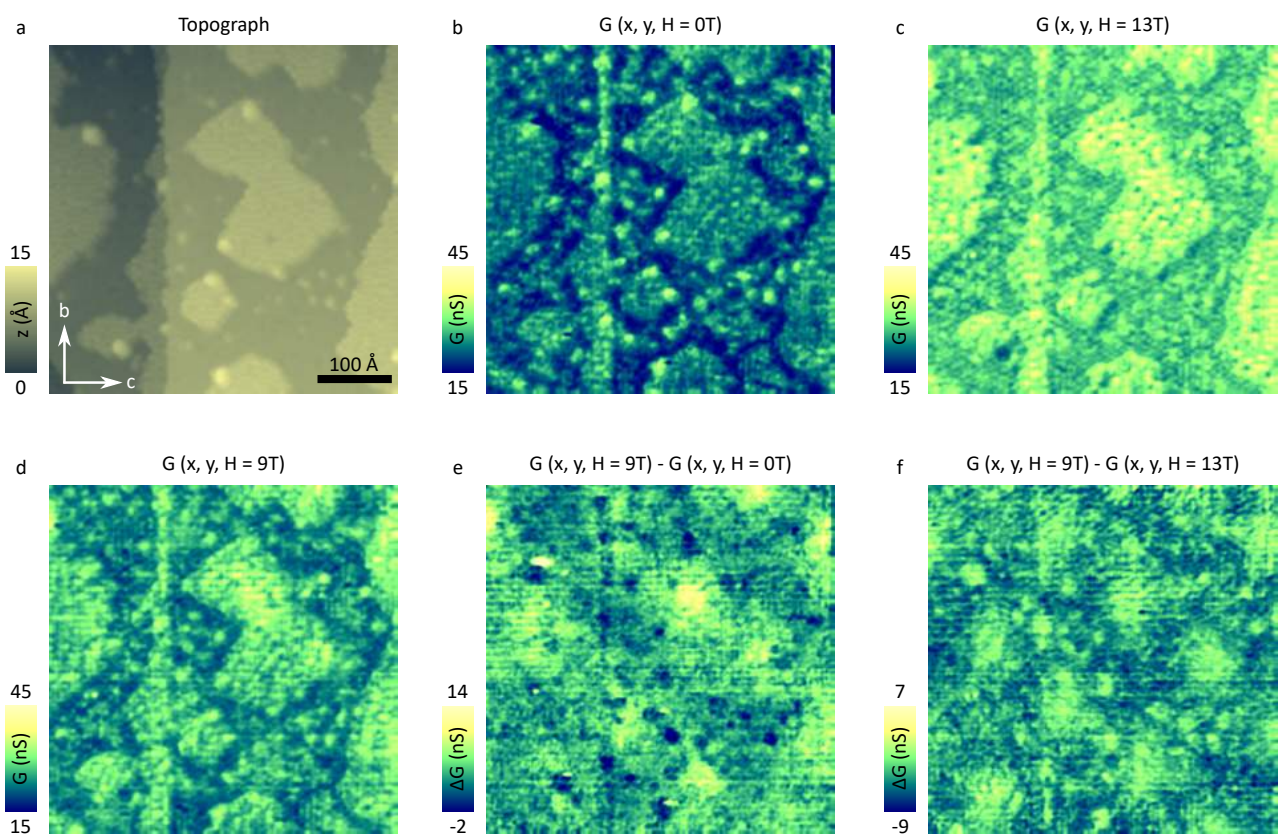

Figure S5

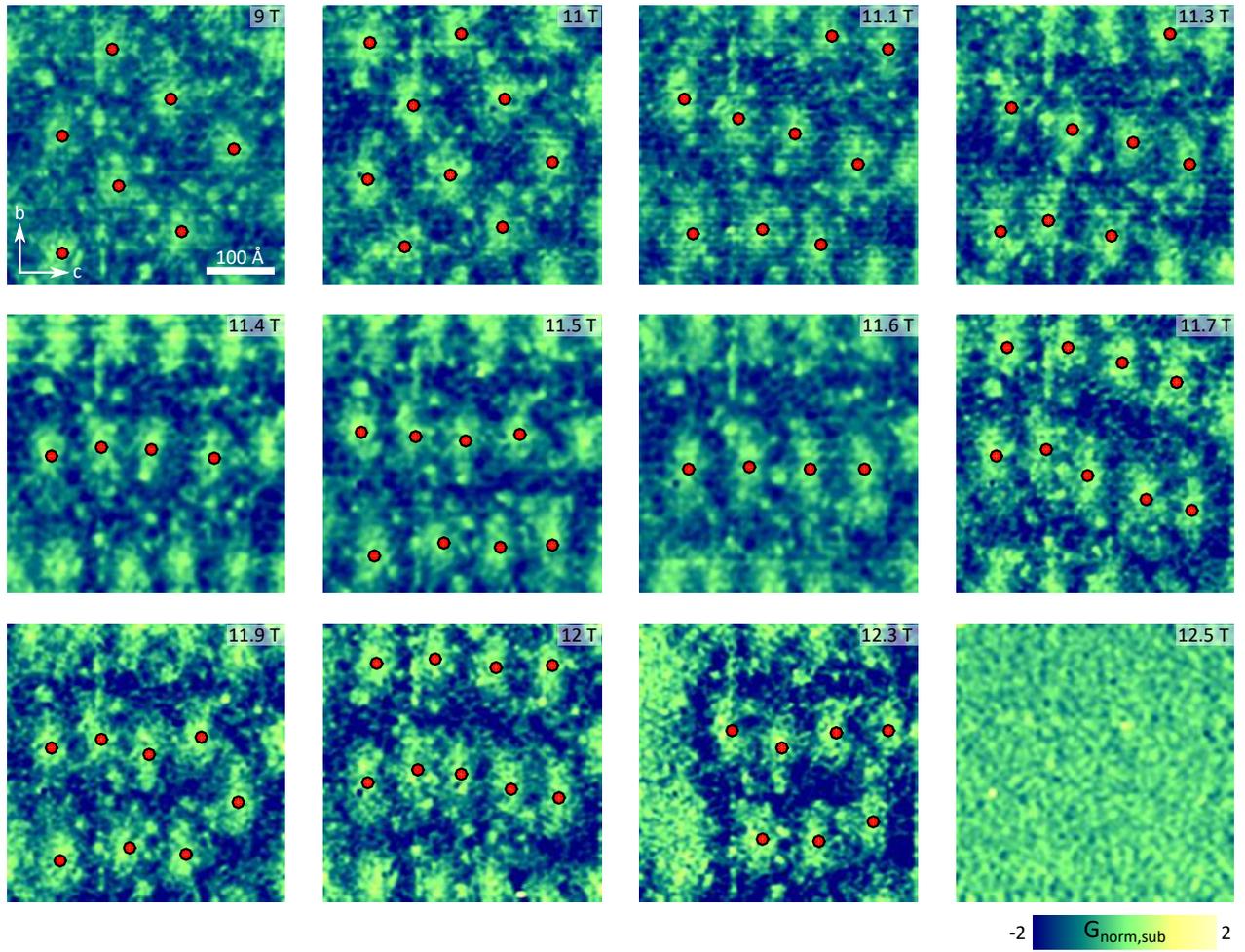

# Figure S6

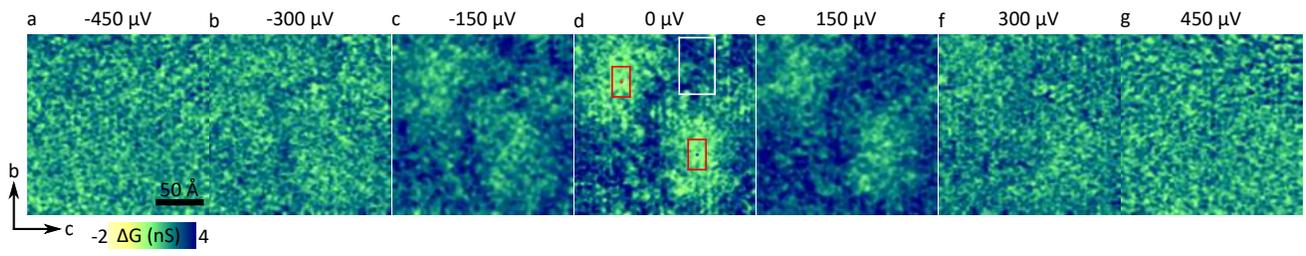